\documentclass[12pt]{JHEP}
\usepackage{amsmath, amsfonts,amssymb,euscript,bbm}
\usepackage{epsfig, cite}
\usepackage{ifthen}
\usepackage{graphicx}
\usepackage{psfrag}

\newboolean{Notes}\setboolean{Notes}{true}
\newcommand{\notes}[1]
	    {\ifthenelse{\boolean{Notes}}{{\tt #1}}{}}

\newcommand{\bas}{\begin{eqnarray*}}
\newcommand{\eas}{\end{eqnarray*}}
\newcommand{\ba}{\begin{eqnarray}}
\newcommand{\ea}{\end{eqnarray}}

\def\ba{\begin{eqnarray}}
\def\ea{\end{eqnarray}}
\def\be{\begin{equation}}
\def\ee{\end{equation}}     
\def\bas{\begin{eqnarray*}}
\def\eas{\end{eqnarray*}}

\def\det{\text{det}}

\def\DD{\text{D}\overline{\text{D}}}

\def\hf{\frac12}
\def\ap{{\alpha^{\prime}}}

\def\p{\partial}
\def\hf{\frac12}

\title{Cosmology of the Tachyon in Brane Inflation}
             
\author{Louis Leblond$^1$
 and Sarah Shandera$^2$\\
  $^1$ George P. \& Cynthia W. Mitchell Institute for Fundamental Physics\\
  Texas A\&M University, College Station, TX 77843-4242\\
\vskip .3cm 
$^2$Institute of Strings, Cosmology and Astroparticle Physics \\
Physics Department, Columbia University, New York, NY 10027\\
\vskip .3cm 
Emails: lleblond@physics.tamu.edu, sarah@phys.columbia.edu}

\abstract{In certain implementations of the brane inflationary paradigm, the exit from inflation occurs when the branes annihilate through tachyon condensation.  We investigate various cosmological effects produced by this tachyonic era. We find that only a very small region of the parameter space (corresponding to slow-roll with tiny inflaton mass) allows for the tachyon to contribute some e-folds to inflation. In addition, non-adiabatic density perturbations are generated at the end of inflation. When the brane is moving relativistically this contribution can be of the same order as fluctuations produced 55 e-folds before the end of inflation. The additional contribution is very nearly scale-invariant and enhances the tensor/scalar ratio. Additional non-gaussianities will also be generated, sharpening current constraints on DBI-type models which already predict a significantly non-gaussian signal.}


\preprint{MIFP-06-29}

\begin{document}

\section{Introduction}
The search for a fundamental realization of inflation, which is a paradigm now well supported by observational data, is most natural in the context of string theory. A particularly attractive idea is brane inflation, where the inflaton is an open string mode describing the brane position in the extra dimensions \cite{Dvali:1998pa}. $\DD$ inflation in type IIB string theory is a realistic implementation of this scenario \cite{Burgess:2001fx, Kachru:2003sx}. There, a mobile $D$-brane moves down a throat toward a $\bar{D}$-brane. The potential comes from the brane interaction and a mass term, and the brane dynamics are given by the DBI action \cite{Silverstein:2003hf, Alishahiha:2004eh}. This model has been particularly interesting because it can be embedded in the flux compactification picture, it is relatively calculable, its predictions can be compared in detail with observations and it has features which distinguish it from many other proposed models \cite{Shandera:2006ax}. Since most observable quantities in simple single-field inflation models depend only on details 50-60 e-folds before the end of inflation, most work has concentrated on model features relevant to this earliest (observable) part of inflation. But one of the attractive features of the $\DD$ model is that inflation has a natural and interesting end. When the branes approach within a string length the tachyon develops and the branes annihilate. This endpoint provides a mechanism for reheating \cite{Barnaby:2004gg} and a possible source for cosmic strings \cite{Sarangi:2002yt}. In general it has been assumed that the details of the annihilation do not have much impact on possible observables other than cosmic string tension. However, previous authors have shown that the tachyon itself may provide an additional period of inflation \cite{Mazumdar:2001mm, Frolov:2002rr, Cremades:2005ir} and that the interactions of the tachyon, the original inflaton and other light degrees of freedom may generate an important contribution to the density perturbation and non-gaussianity \cite{Lyth:2006nx}. Each of these issues has received previous attention in various related contexts, as we will review below. However, here we would like to treat these questions in light of what we have learned about a successful (that is, consistent with known string theory and observational constraints) implementation of the $\DD$ scenario. 

Our primary tool in examining the end of inflation is an action which consistently combines the tachyon T and the primary inflaton $\phi$, related to the brane position. The action for the tachyon itself is a conjecture based on achieving the expected dust behavior when the branes have completely annihilated \cite{Sen:2002in}. However, there is good evidence that this conjectured form is a useful one, and so we will require it as the limit of the combined action when the brane separation is small compared to the red-shifted string scale. We also require the expected DBI action for the brane and anti-brane when the tachyon is very massive, and that the correct mass of the tachyon is recovered.
We discuss in detail in the appendix how one goes smoothly between the two potentials, including details of the Coulombic brane interaction. Here we simply note a few important features. First, both fields have non-standard square root kinetic terms that lead to non-trivial sound speeds \cite{Silverstein:2003hf, Alishahiha:2004eh}. For the tachyon action, the potential multiplies the kinetic term, which leads to very different inflationary dynamics compared to that of the brane position $\phi$ (where the potential is added).  Finally, the terms coupling the tachyon and the brane position are important for generating additional density perturbations. 

With the full action in hand, we address the question of whether a significant number of e-folds may be generated by the tachyon, in addition to those obtained from the usual inflaton. Naively, one might think that the warped throat and square root in the action will lead to a flattening of the potential and to a speed limit for the tachyon, which would enhance the number of e-foldings (as in the brane position case). However, the tachyon only has an inflationary equation of state when it is non-relativistic, and the warping actually makes it {\it more} difficult to get inflation from the tachyon. We find very strong constraints on the parameter space in order for the tachyon dynamics to be important for producing inflation. First, one needs to be in the slow roll regime, which is already arguably fine-tuned because it requires a small inflaton mass. Second, the string coupling $g_s$ must be quite small. This is easy to understand since the height of the tachyon potential is determined by the tension of the D3-branes, which is inversely proportional to $g_s$.  Making $g_s$ very small increases the number of e-folds obtained from the tachyon. On the other hand, a very small $g_s$ also increases the tension of every brane in the model, including the D1-branes that will appear as cosmic strings.  Assuming that these cosmic strings are stable (which requires a mechanism to give a mass to their 2-form coupling \cite{Leblond:2004uc} as well as the absence of monopoles to break upon \cite{Polchinski:2005bg}), we get a lower bound on $g_s$ from the current experimental bound on the cosmic string tension. In summary, the tachyon will contribute to the inflationary dynamics in terms of e-folds in a very small (arguably fine-tuned) region of parameter space, with $g_s$ bounded from below and from above.

From this one might be tempted to say, as was assumed in the past literature, that the tachyonic era is unimportant as far as observable quantities are concerned. Not only are very few e-folds (if any) generated by the tachyon, but they would also be the last e-folds of inflation. The end of inflation is usually considered unobservable since the scales exiting the horizon then are small and re-enter soon after inflation. While this is true for single-field models, there has been much recent work on scenarios that can lead to significant observable, scale invariant density perturbations generated at the end of inflation. These models generically involve the presence of new light scalar degrees of freedom, which lead to non-adiabatic fluctuations that are relevant even for scales that have already left the horizon. A currently popular example is the curvaton model \cite{Linde:1996gt} where an extra light scalar field dominates the energy density at the end of inflation and generates perturbations. Inhomogeneous reheating \cite{Dvali:2003em} can have the same effect and related work on tachyonic preheating can be found in \cite{Barnaby:2006cq}.  In this paper, we examine a related mechanism introduced separately by \cite{Lyth:2005qk} and \cite{Bernardeau:2004zz}. Here inflation ends suddenly due to a previously irrelevant light field that controls the final value of the inflaton, which in turn controls the density.  Because this extra field fluctuates, inflation ends with different densities at different places, which contributes to the density perturbations even on large scales. In a recent paper, Lyth and Riotto \cite{Lyth:2006nx} argue that this mechanism actually happens in brane inflation where the end of inflation is triggered by the tachyon rolling. Taking the end of inflation to occur suddenly at the transition between positive and negative mass for the tachyon, they find that the value of $\phi$ at the end of inflation is a function of other scalar fields and that density perturbations are generated.  They argue that this contribution can be important for certain choice of parameters but they use the standard quadratic derivative for the inflaton which is only good for very small $\dot\phi$ (slow-roll).

In this paper, we show that density perturbations are indeed generated at the end of brane/anti-brane inflation. In the relativistic regime, they can account for at most half the total density perturbations. On the other hand, these density perturbations are negligible in the slow roll regime (contrary to \cite{Lyth:2006nx}). This mechanism also contributes to the amount of non-gaussianity, as discussed for the slow-roll case in \cite{Lyth:2005fi}. Since DBI inflation can saturate the current bounds on non-gaussianity from the CMB, any additional contribution further limits the acceptable string theory parameter space. The distinctive shape of the bispectrum \cite{Chen:2006nt} may also be affected, although a full calculation is not our goal here.

This paper is divided as follows. We first review relevant aspects of the $\DD$ scenario and the simplest actions for the tachyon and inflaton in $\S$2. In $\S$3 we will show that it is generically quite difficult to get significant additional inflation from the tachyon, and the warping of the throat hinders rather than helps the case. $\S$4 deals with the generation of an extra contribution to the density perturbations at the end of inflation, which is particularly interesting when the brane is relativistic at the end. In $\S$5 we discuss the impact of these results on the string theory parameter space and on the expectations for non-gaussianities and we conclude. We leave the details of the geometry and the full action for the inflaton and tachyon to the appendix. Although these details are not necessary to understand our main results, any further calculations of multiple-field effects in brane inflation will need this complete description to achieve the accuracy now found in the data.

\section{The Inflaton and the Tachyon}
Although there are many scalar fields in string theory that may give rise to inflation, most models do not easily achieve enough e-folds (effectively the supergravity ``$\eta$-problem") and do not provide a simple mechanism to end inflation and reheat. In contrast, $\DD$ brane inflation easily provides at least 60 e-folds, agrees with current observations, and the brane collision ends inflation and transfers the inflationary energy to other fields. In this section we will briefly review the key features of brane inflation and the regions of string theory parameter space that fit the cosmological data. The details of the brane collision (described by a tachyon) are usually treated as irrelevant for inflation, but crucial for reheating. However, our goal here is to investigate that assumption, so we also review a few basic features of tachyon cosmology. We will combine the two descriptions in $\S$3.

\subsection{The $\DD$ Scenario}
In brane inflation, we take as much input as possible from known, consistent IIB string theory compactifications and examine the brane interaction in that background. The choice of the brane position, an open string mode, as the inflaton specifies that the dynamics are given by the DBI action. The potential for the brane position is a result of the background geometry and effects coming from moduli stabilization. The geometry is taken to be the simplest calculable example of a smooth warped space - the warped deformed conifold.

With these choices, we have a model with 4 fundamental background parameters: the string scale $m_s=1/\sqrt{\ap}$ and string coupling $g_s$, and the size and cutoff of the warped throat characterized by $N_A$ (the D3 charge) and $h_A$ (the maximum warping) respectively. The brane position has six components, so that the potential has a term $\frac{1}{2}\sum_{i=1}^6m_i^2(\phi^{(i)})^2$. The mass is in principal calculable, but in practice we take it as another (6) parameter(s) which ultimately depends on details of the bulk compactification. In general, we do not expect all of the $m_i$ to be equal and some may be nearly zero thanks to symmetries. We will discuss this point in some detail in $\S$4 and the appendix, but for now we can summarize the relevant equations. We first consider the simple case where only one component of $\phi$ has a significant mass and where the background fields are trivial (except for the contribution of $C_4$ to the Chern-Simons term). Then the action is:
\be\label{DBIact}
S=-\int d^4x\;a^3(t)\left[\mathcal{T}\sqrt{1- \dot{\phi}^2/\mathcal{T}} + V(\phi) - \mathcal{T} \right]\, ,
\ee
where $\mathcal{T}(\phi) = \tau_{3}h^4(\phi)$ is the warped $D$3-brane tension at $\phi$ ($\tau_3 = \frac{m_s^4}{(2\pi)^3g_s}$). The numerical results we will quote for the observational constraints on the string parameter space have used
\be
ds^2=h^{2}(r)\eta_{\mu \nu} dx^\mu dx^\nu + h^{-2}(r)(dr^2+r^2
ds_{5}^2)\, ,
\label{10dmetric}
\ee
with $h(r)\sim r/R$ and $\phi=\sqrt{\tau_3}r$. This is the AdS approximation to the conifold, valid in the large $r$ limit, where the scale of the throat is $R^4=27\pi g_s\ap^2 N_A/4$ and the cutoff is at $h(\phi_A)=h_A$. An investigation of the importance of the full metric was done in \cite{Kecskemeti:2006cg, Shiu:2006kj}. The anti-brane sits at $\phi_A$. In this scenario the potential is 
\be
\label{onephipot}
V(\phi)=\frac{1}{2}m^2\phi^2+2\mathcal{T}\left(1-\frac{1}{N_A}\frac{\phi_A^4}{\phi^4}\right)+\dots
\ee
The interaction term comes from the attraction between the mobile $D$3-brane and an anti-$D$3-brane, and this expression is valid when the $D$3 is not too close to the tip. In principle it is modified for rapidly moving branes, but in practice that correction is not numerically very significant. Generically, we expect $m\sim H$ during inflation \cite{Kachru:2003sx}.

For small $m/M_p$ above, which may be considered a fine-tuning, the scenario is the usual slow-roll inflation. For larger $m/M_p$, the DBI action still allows for many e-folds thanks to the square root in the kinetic term \cite{Silverstein:2003hf, Alishahiha:2004eh}. This defines a local speed of light and so a local speed limit that varies strongly with $\phi$. The speed limit prevents the inflaton from rolling too quickly even along a steep potential. (Note that we will still require the potential energy to dominate the kinetic energy, generically easy for large $m/M_p$.) The importance of this effect can be characterized by a Lorentz-type parameter $\gamma$, where
\be
\label{gammabound}
\gamma = \frac{1}{\sqrt{1- \dot{\phi}^2/\mathcal{T}(\phi)}} \quad \Rightarrow \quad \dot{\phi}^2<\mathcal{T}(\phi)\, .
\ee
Defining pressure and energy density in the usual way, and using the Friedmann equations and the continuity equation gives several useful relations:
\ba
\label{solve}
V(\phi)&=&3M_p^2H(\phi)^2-\mathcal{T}(\gamma(\phi)-1)\, ,\\\nonumber
\gamma(\phi) &=& \sqrt{1+4M_p^4\mathcal{T}^{-1}H^{\prime}(\phi)^2}\, ,\\\nonumber
\dot{\phi}(\phi)&=&\frac{-2M_p^2H^{\prime}(\phi)}{\gamma(\phi)}\, .
\ea
The first two may be used to solve numerically for the Hubble parameter for any value of the input parameters. The third equation is not independent, but will be most useful for us in this paper. 

\subsection{Cosmological Parameters}\label{numbers}
Here we review the basic numerical picture found in \cite{Shandera:2006ax}. These results provide an important starting point for our discussion of the relevance of the tachyon. All cosmological parameters may be expressed in terms of the expansion (flow) parameters
\ba
\epsilon_{D}&\equiv&\frac{2M_p^2}{\gamma}\left(\frac{H^{\prime}(\phi)}{H(\phi)}\right)^2\, ,\\\nonumber
\eta_{D}&\equiv&\frac{2M_p^2}{\gamma}\left(\frac{H^{\prime\prime}(\phi)}{H(\phi)}\right)\, ,\\\nonumber
\kappa_{D}&\equiv&\frac{2M_p^2}{\gamma}\left(\frac{H^{\prime}}{H}\frac{\gamma^{\prime}}{\gamma}\right)\, .
\ea
These are small without the requirement of a flat potential, and $\epsilon=1$ defines the end of inflation since
\be
\label{0eps1}
\frac{\ddot{a}}{a}=H^2(1-\epsilon)\, .
\ee
To first order in these parameters, the scalar spectral index is
\ba
\label{nsminus1}
n_{s}-1 &\equiv& \frac{d\:\ln{\mathcal P}_\zeta}{d\:\ln\:k}\, , \\\nonumber
&\approx& \frac{-4\epsilon+2\eta-2\kappa}{1-\epsilon-\kappa}\, .
\ea 
This becomes exactly zero in the large $\gamma$ (ultra-relativistic) limit. The tensor index is
\ba
n_t &\equiv&\frac{d\ln\:{\mathcal P}_h}{d\ln\:k}\, ,\\\nonumber
&\approx&\frac{-2\epsilon}{1-\epsilon-\kappa}\, ,
\ea
which is non-vanishing even in the ultra-relativistic case. The ratio of power in tensor modes versus scalar modes is
\be
\label{rDBI}
r=\frac{16\epsilon}{\gamma}\, .
\ee
The non-gaussianity parameter $f_{NL}$ is simply proportional to $\gamma^2$, and the detailed bispectrum has been worked out in \cite{Chen:2006nt}.

There are essentially three different regimes of parameter space that fit the data for this model, meaning that they achieve $n_s-1\sim0.95$, $r<0.3$, $|f_{NL}|\lesssim 100$ and match the COBE normalization at 55 e-folds. For the analysis below, two parameters were fixed to simplify searching the parameter space, $g_s=1/10$, $m_s/M_p\sim10^{-2}$, and all 55 e-folds were obtained in the throat. For the purposes of our results, only order of magnitude estimates of the parameter values will be needed. The reader interested in more precise results should refer to \cite{Shandera:2006ax, WMAPcompare}.
\begin{itemize}
\item {\bf{Slow roll}} If $m/M_p\lesssim10^{-10}$, the model behaves exactly as a slow-roll model throughout inflation, with the addition of cosmic strings. This means that $\gamma=1$ to a very good approximation even when the brane is in the bottom of the throat. The dependence on the background parameters simplifies in this case, as discussed in detail in \cite{Firouzjahi:2005dh}. Typical values of the throat parameters are $10^3<N_A<10^6$, $10^{-4}<h_A<10^{-2}$.
\item{\bf{Intermediate}} For $m/M_p\sim10^{-5}$, the scenario may have $\gamma=1$ at 55 e-folds, so that the predictions for the power spectrum look like slow-roll, but the DBI effect is important for obtaining enough e-folds. In this case the bound on $r$ may be saturated and $\gamma$ at the end of inflation can be as large as $10^3$. In this case $N_A\sim10^{14}$ and $h_A\sim10^{-2}$. For large $N_A$, it can be difficult to meet the requirement that the throat be smaller than the bulk \cite{Baumann:2006cd}. We do not worry about that issue here.
\item{\bf{Ultra-relativistic}} For $m/M_p\sim10^{-5}$, $\gamma$ may be large even at 55 e-folds. In this case $r$ is small but the non-gaussianity is large. This is a small shift in background parameters from the previous case, still roughly $N_A\sim10^{14}$, $h_A\sim10^{-2}$.
\end{itemize}
There is an alternative, related scenario where a brane may provide inflation while moving out of the throat \cite{Chen:2004gc}. This scenario may end when the brane subsequently falls down a different throat to annihilate with an anti-brane, in which case our results would apply. Many of these scenarios are reviewed in \cite{Tye:2006uv}.

\subsection{The Tachyon and the Exit of Inflation.}

The exit of inflation in the brane inflationary paradigm is usually triggered by the appearance of a tachyonic mode in the spectrum that signals an instability in the vacuum.  A phase transition follows that ends inflation and starts the Standard Big-Bang cosmology.  This is analogous to hybrid inflation where a ``waterfall" field accomplishes a similar role.  Nevertheless, tachyon condensation in string theory is qualitatively different than usual tachyon condensation in quantum field theory and it is important to remember these differences (there are many reviews on the subject of tachyon condensation in string theory, for example \cite{Sen:2004nf, Ohmori:2001am}).

First, the tachyon potential has a gaussian shape with its VEV at infinity.  The tachyon never reaches its VEV, instead it keeps rolling at the (local) speed of light \cite{Sen:2002nu}.  The final state describes an isotropic fluid with vanishing pressure and constant energy density dubbed ``Tachyon matter" \cite{Sen:2002in}.  

This unusual behavior can be described by effective field theories and there are two actions that have been proposed in the literature to describe this phenomena. First, Sen has proposed a DBI type of action \cite{Sen:2002in} for the tachyon with the potential multiplying the square root terms as follows
\begin{align}\label{tachaact}
S & =-\int d^4x\; a^3(t)AV(T)\sqrt{1-B\dot{T}^2}\; ,\\
V(T)& = \frac{1}{\cosh(\sqrt{\frac{1}{2\ap}}T)}\nonumber\, ,
\end{align}
where $A$ and $B$ are some (geometry dependent) constants and $AV(T)$ is the tachyon potential, so $A$ has dimension four and $B$ is dimensionless. The tachyon T is taken to be real in the equation above and this action is technically valid only for non-BPS branes. However, as shown in the appendix, the full brane/anti-brane action has exactly the same form. This action correctly reproduces the tachyon matter equation of state at late time.  The fact that the potential multiplies the kinetic term is a crucial difference compared to usual DBI inflation as we will see later.  

Another action can be obtained using Boundary String Field Theory \cite{Kraus:2000nj,Takayanagi:2000rz,Jones:2002si}.  This action gives the same physical answer that the tachyon at late time reaches a constant velocity with a constant energy. There are a few differences  \cite{Minahan:2002if}, most importantly that with the BSFT action the pressure goes from negative to positive and then back to zero. In DBI case the pressure is always $\leq0$. This difference could be important for inflation and it is an interesting question to study both actions to see how much their cosmological behavior differs. We leave this problem for future work and in the rest of this paper we will use Sen's effective action because it is simpler.

Now, the tachyon might be able to give inflation, and many authors have investigated the question of how many e-folds can be obtained in various scenarios. Usually, the potential for the tachyon is too steep for slow-roll inflation.  Assisted inflation \cite{Mazumdar:2001mm} can fix this problem if there are multiple tachyon fields: the large Hubble friction effectively slows down the tachyons irrespective of the steepness of their potential. Here, there is a single tachyon field and so no help from Hubble friction.  Nevertheless, tachyon condensation happens at the bottom of a warped throat and one may wonder if the speed limit can lead to a DBI type of inflation for the tachyon.  We will show that this is not the case.

\subsection{Warped Tachyon Cosmology}
Here we review a few basic expressions for tachyon cosmology in a warped geometry \cite{Raeymaekers:2004cu, Cremades:2005ir}. We start with a simple action, assuming that only the tachyon field is relevant. This model demonstrates a few properties of the tachyon action that we will use later.
\be
S=-\int d^4x\; a^3(t)AV(T)\sqrt{1-B\dot{T}^2}\, ,
\label{tachact}
\ee
where $A$ and $B$ are some (geometry dependent) constants and $AV(T)$ is the tachyon potential. 
At the bottom of the throat (at $\phi_A$), where the D3-brane and the anti D3-brane meets, $A = 2\tau_3 h_A^4$ and $B = h^{-2}_A$.
Then the pressure $\rho_T$ and density $p_T$ are given by
\ba
\rho_T&=&AV(T)\gamma_T\, ,\\\nonumber
p_T&=&-\frac{AV(T)}{\gamma_T}\, ,
\label{tachprho}
\ea
where we have defined the tachyon Lorentz factor $\gamma_T$ by
\be
\gamma_T=\frac{1}{\sqrt{1-B\dot{T}^2}}\, .
\ee
From the continuity equation
\be
\label{Tdot}
\dot{T}=\frac{-2M_p^2H^{\prime}}{\gamma_T\sqrt{ABV}}\, ,
\ee
where primes denote derivatives with respect to the canonical field. Notice that for the tachyon to be the source of inflation, we need
\ba
\label{Tinf}
\rho_T+3p_T&<&0\, ,\\\nonumber
\Rightarrow\dot{T}^2&<&\frac{2}{3B}\, .
\ea
The speed limit coming from the square root in the action only requires
\be
\dot{T}^2<\frac{1}{B}\, ,
\ee
so if the tachyon is moving too fast (large $\gamma_T$), there is no inflationary phase. If the tachyon does contribute, one must use the canonical field
\be
\phi_T=\sqrt{AB}\int \sqrt{V(T)} dT
\ee
Observables can be computed using the relationship between derivatives:
\be
\frac{d}{d\phi_T}=\frac{1}{\sqrt{ABV(T)}}\frac{d}{dT}.
\ee

\section{Tachyon Cosmology}

\subsection{The Sudden End Approximation}
We will assume that the time during which neither the original inflaton nor the tachyon dominates the dynamics is short. We will find that this is justified by examining the value of $T$ when the tachyon dominates energetically. To examine the effect of the tachyon, we require energy conservation, so that all of the energy density in the inflaton when the tachyon develops (when the brane separation is the warped string length) must be in the tachyon field when $\phi$ is negligible. The energy densities for the separate fields are, using (\ref{DBIact}, \ref{Tinf}),
\begin{align}
\label{densities}
\rho_\phi & = \tau_3 h_A^4 (\gamma_A -1) + V(\phi_A)\, ,\\
\rho_T & = 2 \tau_3h_A^4 V(T) \gamma_T\, .\nonumber
\end{align}
There are two cases, corresponding to relativistic and non-relativistic brane motion. In either case, it is reasonable to assume the energy transfer happens quickly. Then the tachyon is a significant fraction of the energy density at $T\approx 0$ and $V(T) \approx 1$. One may check directly that in the slow-roll case this makes sense: the height of the tachyon potential is $V_0$, so as soon as it starts rolling it has an energy density comparable to that in the inflaton. The same is true in the DBI case in the AdS approximation. In inflation with the original inflaton it is assumed that the potential energy dominates the kinetic energy. This is reasonable because the slow-roll case has $\gamma\approx1$ and $V(\phi)\approx2\tau_3h_A^4=V_0$, while the DBI case has large $\gamma$ but $V(\phi)\approx m^2\phi^2/2\gg V_0$. From Eq.(\ref{densities}), if the inflaton is rolling slowly at the end of inflation then $\rho_T=2\tau_3h_A^4$, so that $\gamma_T\approx1$. On the other hand, if $V(\phi_A)\gg V_0$ then $\gamma_T$ must be large.  Since $\dot T$ must be small to get inflation, the tachyon can only contribute to inflation in the slow roll regime where $\gamma \sim 1$. Of course, it could be that the inflaton itself continues to generate a significant number of e-folds after the tachyon starts rolling. However, one can check that even ignoring the tachyon, the number of e-folds from the inflaton between $\phi_e=h_Al_s + \phi_A$ and $\phi_A$ is less than one. That calculation assumes the AdS metric and parameter values that already match the COBE normalization. It may need to be re-evaluated using the full metric. 

\subsection{Estimating Tachyon E-folds}
Based on the discussion in the previous section, we may use the standard slow-roll analysis to determine the number of e-folds from the tachyon, assuming $\gamma_T$ starts out very close to 1. This gives
\ba
\label{efoldsTach}
N_{e,T}&=&-\frac{AB}{M_p^2}\int_{T_i}^{T_f}\frac{V^2}{V_{,T}}dT\, ,\\\nonumber
&=&\frac{2\tau_3h_A^2}{M_p^2}\int_{T_i}^{T_f}\frac{1}{\sqrt{\frac{1}{2\ap}}\sinh{\left(\sqrt{\frac{1}{2\ap}} T\right)}}dT\
\ea
The final value of $T$ can be determined using the condition for inflation, Eq.(\ref{Tinf}), and Eq.(\ref{Tdot}). Then
\be
\frac{\left[\sinh{\left(\sqrt{\frac{1}{2\ap}} T_f\right)}\right]^2}{\cosh\left({\sqrt{\frac{1}{2\ap}} T_f}\right)}=\frac{4\tau_3h_A^2}{(\frac{1}{2\ap})M_p^2}\, .
\ee 
The right hand side is generically less than 1, so that the solution for the expression above is at $\sqrt{\frac{1}{2\ap}} T_f<1$, and we may expand the left hand side. Doing so gives
\be
T_f\approx\frac{4h_A}{(2\pi)^{3/2} g_s^{1/2}}\frac{1}{M_p}\, .
\ee
Conservation of energy can be used to determine the initial value of $T$, as described in the previous subsection:
\be
2\tau_3h_A^4+\frac{1}{2}m^2\phi_A^2=2\tau_3h_A^4V(T_i)\gamma_{T_i}\, .
\ee
Expanding both $V(T)$ and $\gamma(T)$ for small $T$, small $\dot{T}$ and keeping only the most important terms gives
\ba
\dot{T_i}^2&\approx&\frac{m^2\phi_A^2}{2\tau_3h_A^2}\, ,\\\nonumber
\Rightarrow T_i&\approx&\frac{m}{m_s}(27\pi g_sN)^{1/4}\sqrt{\frac{3}{4\pi^3 g_s}}\frac{1}{M_p}\, ,
\ea
where we have used Eq.(\ref{Tdot}) and the slow-roll relationship between $H(\phi)$ and $V(\phi)$. Then, the condition that the tachyon does not initially move too quickly to provide inflation is the condition $T_i<T_f$. This translates to a fairly strong condition on the background parameters:
\be
\frac{m}{M_p}<\left(\frac{m_s}{M_p}\right)\frac{4h_A}{(\sqrt{6})}\frac{1}{(27\pi g_sN_A)^{1/4}}\, .
\ee
This condition can only be met for typical values of the parameters in the slow roll regime, because of the tension between $N_A$ and $m$. This is consistent with our assumption that $\gamma$ is small.

Returning to the calculation of the number of e-folds due to the tachyon and performing the integral in Eq.(\ref{efoldsTach}) (expanding for small $T$) gives
\be
\label{efoldsT2}
N_{e,T}=\frac{4h_A^2}{(2\pi)^3g_s}\left(\frac{m_s}{M_p}\right)^2\ln\left[\frac{T_f}{T_i}\right]\, .
\ee
This can only be large if $g_s$ is small, or if $T_f\gg T_i$ which requires $m/M_p$ to be extremely small. Decreasing $g_s$ drives the brane tension up, which also increases the cosmic string tension. In fact, there is effectively a bound on $g_s$ from the string tension using the slow-roll expression derived in \cite{Firouzjahi:2005dh}:
\be
G\mu=\left(\frac{3\times5^6\times\pi^2}{2^{21}}\right)^{1/4}g_s^{-1/2}\delta_H^{3/2}N_e^{-5/4}f(\beta)\, ,
\ee
where $\delta_H$ is the COBE normalization and $\beta=(3/2)(2\pi)^3 h_A^{-4}(M_p/m_s)^4(m/M_p)^2g_s$. The function $f(\beta)\rightarrow1$ for $\beta\rightarrow0$, which happens for small mass and small $g_s$. Using $N_e=55$ and $\delta_H=10^{-5}$, the bound on the cosmic string tension $G\mu\lesssim10^{-7}$ \cite{Pogosian:2003mz} implies $10^{-6}\lesssim g_s$. These calculations show that generically it is very difficult to generate a significant number of e-folds with the tachyon. It is interesting that the warping of the throat actually makes it more difficult to get inflation here (Eq.(\ref{efoldsT2}) is suppressed by $h_A^2$) in contrast with brane inflaton, because of the very different equation of state for the tachyon.

\section{Density Perturbations at the End of Brane Inflation}

\subsection{Multiple Fields in Brane Inflation}
Brane inflation has so far been treated as a model with a single scalar field. For most of the inflationary period, this works. Although the brane position is always a six-component field, we generally take the inflaton to be the linear combination that actually controls inflation. For slow-roll, where the mass term is small, that direction is the one perpendicular to the branes. When the mass term is large, the important field is whichever combination appears in the K\"{a}hler potential and superpotential. This is arguably the less fine-tuned scenario. However, it is reasonable to expect that there will be some (approximate) symmetries that protect some components of the inflaton from getting a large mass. The essential point for our results below is simply that the potential generically depends differently on different components of the inflaton, while the condition for the development of the tachyon depends on the separation between the branes. We give further justification for this in the appendix.

As an example, we can look at the $\DD$ action at the bottom of the throat
 including the tachyon $T$, the inflaton $\phi$ (taken to be 
the radial position of the D3 brane) and an extra scalar field $\sigma$ taken to be the position of the brane on a circle $S^1$ (see the appendix for a complete analysis).  This is a simplified picture since, at the bottom of the throat, the geometry is really that of an $S^3$ and one should have three angles.
\begin{align}\label{smallphi}
S = &- \int d^4 x V(T)V^\prime(T,\phi,\sigma)\tau_3h_A^4 a^3 \left(\sqrt{1 - \dot\phi^2h_A^{-4} \tau_3^{-1}- \dot\sigma^2h_A^{-4}\tau_3^{-1} - h_A^{-2} \dot T^2} \right.\\
& \left. + \sqrt{1 - h_A^{-2} \dot T^2}\right) + V(\phi,\sigma)\nonumber \, ,
\end{align}
where the potential $V(T)V^\prime(T, \phi, \sigma)$ is 
\begin{align}\label{tachyonmass}
V(T)V^\prime(T,\phi, \sigma) & = \frac{1}{\cosh(\sqrt{\frac{1}{2\ap}}T)}\sqrt{1+ \frac{((\phi-\phi_A)^2 +\sigma^2) h_A^{-2}\tau_3^{-1}T^2}{(2\pi\ap)^2}}\nonumber\\
&  \approx  1 + \frac{T^2}{2} \left( -\frac{1}{2\ap} + \frac{((\phi - \phi_A)^2 +\sigma^2) \tau_3^{-1}h_A^{-2}}{4\pi^2\ap^2}\right) + \mathcal{O}(T^4)\, ,
\end{align}
and $V(\phi,\sigma)$ contains the coulombic and quadratic terms.  Note that as the brane separation goes to zero the coulombic part should disappear from the potential and one should only be left with the mass term and the tachyon potential (see appendix for details). Since here we consider the action for separations close to the warped string scale (and slightly greater) we keep the coulombic piece. 

The constant piece of the potential in (\ref{onephipot}) ($V_0 = 2\tau_3h_A^4$) is coming from the second square root (with $\dot T \sim 0$) as well as from the factor of one in the first kinetic term when we expand for small $\dot \phi$. For small $\phi$, the Chern-Simon part of the action (\ref{smallphi}) is essentially zero due to the exact cancellation between the D3 and $\overline{D3}$, but for large $\phi$ there are two terms and the anti-D3 brane part contributes to the constant part of the potential.  All in all, it is quite intricate and interesting how one can write a general action that incorporates both the tachyon and the inflaton in a consistent manner.

The quadratic part of the potential is $V(\phi, \sigma) = \hf m^2_\phi \phi^2 + \hf m^2_\sigma \sigma^2 + \cdots$ and under the assumption that the $\sigma$ direction represents an approximate isometry of the Calabi-Yau, we have that $m_\sigma \ll m_\phi$. We then take $\phi$ to be the inflaton and $\sigma$ to be a spectator field.  The end of inflation happens when the tachyon starts rolling\footnote{It is possible for inflation to end before that through tunneling (see \cite{Jones:2002si} for example), in this work we neglect this aspect as it is exponentially suppressed and in most regions of spacetime, inflation ends through tachyon rolling.}. In the spirit of the sudden end approximation used in the previous sections, we can consider that this happens suddenly when the mass of the tachyon goes from positive to negative in (\ref{tachyonmass}).  The value of $\phi$ at the end of inflation is $\phi_e(\sigma) = \phi_A+\sqrt{2\pi^2\ap\tau_3 h_A^2 - \sigma^2}$.

Because of the presence of the $\sigma$ term, inflation will not end on a constant-density surface and perturbations on all scales, even those already outside the horizon, will receive a second contribution proportional to that density difference. In the DBI scenarios the non-standard kinetic term generates a large $\gamma$ factor and allows the inflaton to generate many e-folds in spite of a steep potential. This same sensitivity to position will exaggerate the difference in number of e-folds obtained for regions with slightly different field values, making this second source of perturbations substantial. In other words, DBI inflation generates density perturbations with a size that is independent of $\phi$ since the growth in $\gamma$ exactly cancels the shrinking Hubble scale. So contributions generated at any point during inflation can potentially be important. The answer is general in the sense that even if the brane is moving slowly at 55 e-fold, it can move quickly at the end of inflation and the large $\gamma$ (DBI) limit can be used to calculate this contribution. It is only in the case where the brane is still moving slowly at the end that one finds a different answer. This would arguably be the most fine tuned scenario and so we can consider the DBI limit as the generic case.

\subsection{The $\delta N$ Formalism}

\begin{figure}[htb]
 \begin{center}
\includegraphics[width=0.7\textwidth,angle=0]{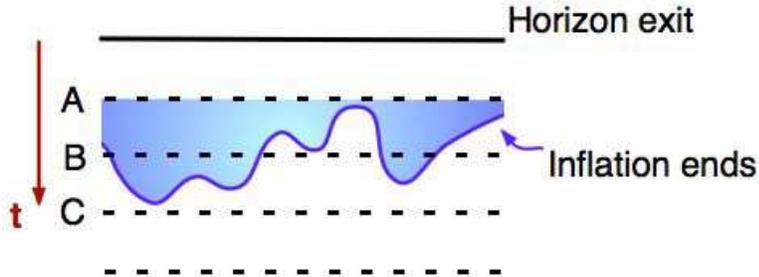} 
\caption{To calculate the density perturbation, we must add the contribution from the end of inflation. The dashed lines represent surfaces of uniform density while the curved line is the surface inflation ends on. The arrow indicates the direction of time. Between horizon exit and A, the superhorizon modes do not evolve. However, to calculate the size of the perturbation that will remain after inflation (line C), we must calculate the contribution from the shaded region.
\label{InfEnds}}
\end{center}
\end{figure}

To examine the production of density perturbation at the end of brane inflation in detail, we use the $\delta N$ formalism for calculating perturbations due to multiple fields \cite{Wands:2000dp, Sasaki:1995aw}. In the $\delta N$ formalism one writes the power spectrum as a sum of contributions from each field. We take the standard definition for the number of e-folds, $N_e$, as
\be
N_e=\int H dt\, .
\ee
From the continuity equation, it is easy to see that uniform density hypersurfaces are separated by uniform expansion for adiabatic perturbations:
\be
\frac{d\rho}{dN_e}=-3(\rho+p)\, ,
\label{adiabatic}
\ee
where $p$ is a unique function of $\rho$, so that this expression could be solved for $N_e(\rho)$. Figure (\ref{InfEnds}) illustrates that once the perturbation exits the horizon, it does not evolve until the second field becomes important, at surface A. Surfaces A, B, and C are uniform density, but they are not separated by uniform expansion because inflation ends at different times between them. From Eq.(\ref{adiabatic}) then, we are dealing with non-adiabatic modes, as expected when multiple fields are involved and the superhorizon perturbation may evolve. We may evaluate the amount it evolves by calculating how many e-folds each region undergoes between A and C, due to the additional field, $\delta N_e$. The result can be carried forward from the time just after inflation to today, remaining fixed until it reenters the horizon.

We will be interested in the perturbation on constant-density hypersurfaces, $\zeta$, which in the single field (or adiabatic) case becomes constant after exiting the horizon. The power spectrum is defined by the Fourier components $\zeta_k$ as
\be
\label{powspect}
\langle\zeta_k\zeta_{k\prime}\rangle=\frac{2\pi^2}{k^3}\mathcal{P}_{\zeta}\delta^3(k-k^{\prime})\, .
\ee
For a generic set of fields $\phi^a$, $\delta N$ will depend on all of them \footnote{$\delta N$ is also a function of  $\dot{\phi}^a$.  In the slow roll regime we can neglect these terms in the expansion.  But even if one does not use slow roll, one can use the equation of motion for $\phi^a$ (\ref{solve}) to eliminate any  dependence on $\dot\phi^a$ in favor of $\phi^a$.}
\be
\label{Nexpansion}
\zeta=\sum_a\frac{\partial N}{\partial\phi^a}\delta\phi^a+\sum_{ab}\frac{\partial^2N}{\partial\phi^a\partial\phi^b}\delta\phi^a\delta\phi^b+\dots
\ee
where
\be
\frac{\partial N}{\partial\phi^a}=-\frac{H}{\dot{\phi^a}}\, ,
\ee
and the perturbations $\delta\phi^a$ must be solutions to the equation of motion for the perturbed scalar fields. The first term clearly recovers the standard slow-roll result for a single field where we take $\langle\delta\phi_k\delta\phi_{k}\rangle \sim \frac{H}{2\pi}$.  Hence the first term reproduces the correct power spectrum to zero order in the slow-roll parameters (or DBI parameters).  One can easily check that $n_s -1$ calculated from this will have exactly the right first order dependence on the DBI parameters (including the $\kappa$ factor in DBI coming from derivative of $\gamma$ \cite{Shandera:2006ax}).  Higher order terms in (\ref{Nexpansion}) will give among other things the bispectrum \cite{Sasaki:1995aw}.

In the case at hand, inflation does not end on a constant-density hypersurface, and so we must patch on that contribution as follows
\ba
\mathcal{P}_{\zeta}&=&\left[\left(\frac{\partial N}{\partial\phi}\right)^2+\left(\frac{\partial N}{\partial\sigma}\right)^2\right]\left(\frac{H}{2\pi}\right)^2\, ,\\\nonumber
&=&\left.\left(\frac{H^2}{2\pi\dot{\phi}}\right)^2\right|_{k=aH\gamma}+\left.\left(\frac{\partial N}{\partial\phi}\right)^2\left(\frac{H}{2\pi}\right)^2\right|_{\phi_{end}}\left(\frac{d\phi_{end}}{d\sigma}\right)^2\, ,
\ea
where
\be
\frac{\partial N}{\partial\phi}=\frac{H}{\dot{\phi}}=\frac{H\gamma}{2M_p^2H^{\prime}}=\frac{1}{\sqrt{\epsilon}}\sqrt{\frac{\gamma}{2M_p^2}}\, .
\ee
The second term can be thought of as coming from the shaded region in Fig. \ref{InfEnds}. Even if $\gamma$ is relatively small at 55 e-folds ($\sim2$, say), it may grow several orders of magnitude by the end of inflation since the warp factor (speed limit) depends so strongly on position. On the other hand $H/H^\prime$ is constant for large $\gamma$ and so $\frac{\partial N}{\partial\phi}$ grows towards the end of inflation.  We expect $d\phi_{end}/d\sigma\sim1$ since the size of the $S^3$ at the bottom of the throat is roughly $\sqrt{g_sM}h_Al_s$. That is, the angular direction is as important as the radial direction for determining the end of inflation. This is illustrated for the warped throat in the appendix. So even if the DBI behavior has only a moderate effect on predictions based on the analysis near 55 e-folds, it may require a recalculation of the size of density perturbations due to effects at the end of inflation.  Note that even though the variation of the number of e-folds grows with respect to $\phi$, the density perturbation $\mathcal{P}_{\zeta}$ also contains an extra factor of $H^2$ which decreases in the DBI limit.

In the preceding discussion, we have supposed that all the inflationary dynamics are dominated by the inflaton $\phi$.  But the conclusion is essentially unchanged if the tachyon or $\sigma$ are important since different regions will still have different densities at the end of inflation. 


\subsection{Numerical Values}

So we have found that brane inflation will receive an additional contribution to the density perturbation at the end of inflation of the order of
\be
\mathcal{P}_{\zeta}\approx\left(\frac{\partial N}{\partial\sigma}\right)^2\left(\frac{H}{2\pi}\right)^2
\approx \frac{H^2\gamma}{8\pi^2M_p^2 \epsilon}\, .
\ee
This effect depends on the competition between H, $\gamma$ and $\epsilon$. 
In the DBI regime where $\gamma \gg 1$, we have that 
\begin{align}
H(\phi) & \sim \frac{ m}{M_p}\frac{\phi}{\sqrt{6}}\, ,\\
\label{largegam} \gamma(\phi) & \sim \frac{m M_p}{\phi^2}\sqrt{\frac{2\lambda}{3}}\, ,\\
 \epsilon & = \sqrt{\frac{6}{\lambda}}\frac{M_p}{ m}\, ,
 \end{align}
where $\lambda = \tau_3 R^4 = \frac{27N_A}{32\pi^2}$.  These expressions are derived using a warp factor $h \propto \phi$. In the full solution $h$ is approximately constant at the bottom of the throat. This should lead to a leveling of $\gamma$ rather than a continued $1/\phi^2$ behavior, so the relations above give an upper bound on the density perturbation at the bottom of the throat. The actual value will be smaller once the deformed metric is accounted for. So if $\gamma$ is large at the end of inflation the additional contribution will be
\begin{align}\label{DBIlimit}
\mathcal{P}_{\zeta}^{\text{end}} \sim \left(\frac{ m}{M_p}\right)^4\frac{\lambda}{18}\sim 10^{-3}\left(\frac{ m}{M_p}\right)^4N_A\, ,
\end{align}
which is independent of the warp factor and $\phi$.  For the DBI regime, this is exactly the same size contribution as the usual one obtained 55-efolds before the end of inflation. Assuming the parameters $m$ and $N_A$ have already been adjusted to match the COBE normalization, $\mathcal{P}_{\zeta}^{\text{end}}\sim10^{-10}$. In the intermediate regime the brane is moving relativistically at the end of inflation, so the size of the contribution depends only on the parameters $m$ and $N_A$ using the DBI expression above. Since the parameter values are quite similar to those for DBI, we again expect ${P}_{\zeta}^{\text{end}} \approx 10^{-10}$.

For the case where $\gamma=1$ to a good approximation at the end of inflation, the result (\ref{DBIlimit}) does not apply.  A similar calculation yields the density perturbation at the end of inflation for the slow-roll regime, where the coulombic term is important \footnote{As discussed in more detail in the appendix, the coulombic potential is actually modified for small $\phi$. We are using it for simplicity but one can in principle calculate the full potential \cite{Jones:2002si}.} ($V \sim V_0 + m^2\phi^2 -\frac{V_0\phi_A^4}{N_A(\phi-\phi_A)^4}$):
\begin{align}\label{SRdensity}
\mathcal{P}_{\zeta}^{\text{end}} &\sim \left.\left(\frac{V}{V^\prime}\right)^2\frac{H^2}{(2\pi)^2} \frac{1}{M_p^4}\right|_{\phi-\phi_A = h_Am_s}\\
\sim & \frac{1}{12(27)^2 g_s^{4}} h_A^6 \left(\frac{m_s}{M_p}\right)^6
\end{align}
 where $m_s = 10^{-2}M_p$ is the string scale and $h_A$ the warp factor at the bottom of the throat. We have replaced $\phi_A$ by its dependence on the warp factor, and since the warp factor is approximately constant at the bottom of the throat we are justified to use $h(\phi_A)$ instead of $h(\phi_e$).  For the range of parameters appropriate for slow-roll, we find a negligible amount of density perturbations at the end of inflation ($\sim 10^{-23}$).  
 
Note that we are justified to consider $\sigma$ as a spectator field since the $m^2\phi^2$ term is still the important non-constant piece. Indeed in slow-roll, the potential is approximately $V\approx V_0$ but the $m^2\phi^2$ is still bigger that the coulombic piece (for most values of $\phi$) and it determines the inflaton direction.  On the other hand, in the derivative of the potential the coulombic contribution dominates over the mass term.  Even though $m^2$ does not appear in (\ref{SRdensity}), it is important in the potential and we are justified to consider $\phi$ as the primary inflaton. However, the fact that the potential is nearly constant already leads us to expect that the effect of the extra fields cannot be large. That is, the energy density at the end of inflation is relatively insensitive to the inflaton direction. 
 
In summary, the density perturbations generated at the end of inflation are negligible for the slow-roll regime but they can account for up to 50\% of the total density perturbation for the intermediate or DBI regime.

\section{Discussion}

In this paper, we have shown that there are effects from the tachyon at the end of brane inflation that may significantly affect the calculation of the observables. We have found that while it is generically very difficult to get much inflation from the tachyon, additional density perturbations will be generated. These are negligible in the slow-roll regime but important when the brane is relativistic at the end of inflation, where they can contribute up to 50\% of the total power. This extra contribution is very nearly scale invariant, since in the DBI limit (with the AdS metric) $n_s=1$ to first order in $\epsilon$. Fortunately, we do not find that this contribution dominates, which would make it difficult to obtain a scalar tilt much different from $n_s\approx1$ (now disfavored by observation).

Adding this contribution will not change the order of magnitude of the background parameters needed to fit the data. However, it will change the prediction for the magnitude of the tensor/scalar ratio given in Eq.(\ref{rDBI}). To see how, note that the addition of the extra piece above means that the background parameters must be adjusted slightly relative to the results of \cite{Shandera:2006ax} to match the measured amplitude of fluctuations. This means decreasing either $m$ or $N_A$ by a small factor. From Eq.(\ref{largegam}), either of these choices decreases $\gamma(\phi)$. The number of e-folds benefits by a factor of $\gamma$ as
\be
N_e=-\frac{1}{2M_p^2}\int_{\phi_i}^{\phi_e}\frac{H(\phi)\gamma(\phi)}{H^{\prime}(\phi)}\:d\phi
\ee
so that decreasing $\gamma(\phi)$ means that inflation must begin at a larger value of $\phi$ to still obtain enough e-folds. The power in tensor modes, which is not altered by the tachyon dynamics, is
\be
\mathcal{P}_h=\frac{2H^2}{M_p^2\pi^2}+\mathcal{O}(\epsilon^2)
\ee
Since $H\sim\phi$, increasing the value of $\phi$ at 55 e-folds ($\phi_{55}$) will increase $H(\phi_{55})$ and so increase the power in tensor modes. Since the power in scalar modes is fixed by the COBE normalization, the ratio $r$ is enhanced. This may loosen the bounds on the DBI parameter space found in \cite{Baumann:2006cd}.

DBI inflation already has quite enhanced non-gaussianty. Since the sound speed ($c_s\sim1/\gamma$) can be much smaller at the end of inflation than at the beginning, any additional contribution may be quite significant. Even an additional factor of 2 in $f_{NL}$ is quite significant because it further limits the value of $\gamma$ allowed at 55 e-folds, which in turn alters the range of other predictions. In \cite{Lyth:2006nx}, the non-gaussianities are calculated assuming a negligible primordial bispectrum and the usual sound speed ($c_s=1$). This assumption is probably not valid in the relativistic regime where the primordial contribution to the density perturbations is expected to be quite important \cite{Alishahiha:2004eh, Chen:2006nt, Chen:2005fe}. In addition, it may be worthwhile to compare the shape of bispectrum generated at the end. Interestingly, \cite{Creminelli:2006gc} shows that the most important shape for non-gaussianities generated this way are the squeezed limits, where one momenta is much smaller than the others. This is also the interesting shape for single field DBI non-gaussianity \cite{Chen:2006nt}.  

Finally, the analysis in this paper assumes that inflation ends due to brane annihilation - an aspect of the scenario that naturally gives rise to reheating and cosmic strings. However, a $D3$ brane can experience a potential even in the absence of the anti-brane and we may still consider its position to be the inflaton \cite{Dymarsky:2005xt}.  In this kind of scenario it is less clear how inflation ends and in particular there is no apparent reason for the end of inflation to depend on extra scalar fields.

As this paper was being finished, some related work on generating density perturbations at the end of brane inflation was worked out in \cite{Matsuda:2006ee, Matsuda:2006ie}\\
 
 \noindent {\bf \large {Acknowledgment}}

We wish to thank Melanie Becker, Aaron Bergman, Lam Hui, Justin Khoury, Gary Shiu, Henry Tye, Justin F. Vazquez-Poritz, Bret Underwood and Mark Wyman for valuable discussions.  We would like to thank the organizers of the Simons Workshop in Mathematics and Physics 2006 where this work was started. L.L. is also grateful to the Perimeter Institute for its hospitality while this work was being finished. The work of L.L.~has been supported by the National Science Foundation under grant PHY-0555575 and the University of Texas A\&M. The work of S.S. is supported by the DOE under DE-FG02-92ER40699.
 
\appendix

\section{Joining the Tachyonic Era to the Inflationary Era}\label{tachyon}
\subsection{Effective Action}
In this section we work out explicitly the coupling between the tachyon and the inflaton field.  We follow the procedure of \cite{Garousi:2000tr} that consists of writing (via some educated guesses) the action for the tachyon for $D9\overline{D9}$ and then applying T-duality to obtain the action for lower dimensional branes. We will also examine the connection between the action usually written down for the inflaton when the branes are far apart and the action appropriate for the tachyon era. In the following we will use the definitions
\begin{align*}
E_{\mu\nu}& = g_{\mu\nu} + B_{\mu\nu}\, ,\\
\lambda &= 2\pi\ap\, .
\end{align*}
Our starting point is therefore the action proposed by Sen \cite{Sen:2003tm} (also proposed in \cite{Garousi:2000tr}) for the brane anti-brane action (for $D9\overline{D9}$) 
\begin{align}\label{tachSen}
S = -\tau_9\int d^{10}x \; V(T) \left(\sqrt{-\det\; {\bf A_1}} + \sqrt{-\det\; {\bf A_2}}\right)\, ,
\end{align}
where 
\begin{align*}
{\bf A^{(i)}_{\mu\nu}} & = E_{\mu\nu} + \lambda F^{(i)}_{\mu\nu} + \hf (\overline{D_\mu T}D_\nu T +\overline{D_\nu T}D_\mu T)\, ,\\
F^{(i)}_{\mu\nu} & = \p_\mu A^{(i)}_\nu -  \p_\nu A^{(i)}_\mu\, , \\
 D_\mu T & = \left(\p_\mu + i(A^1_\mu - A^2_\mu)\right)T\, ,\\
 V(T)& = \frac{1}{\cosh(\sqrt{\frac{1}{2\ap}}T)}\nonumber\, .
\end{align*}
 Note that for a vanishing gauge field and for a real tachyon, this action does reproduces (\ref{tachaact}). This action also reproduces the tachyon matter dust expected from string field theory \cite{Sen:2003tm}. It includes consistently the gauge field such that on top of the tachyon potential (at $T=0$) the action reduces to a $U(1)\times U(1)$ DBI action for brane anti-brane.  In this picture, the tachyon is a string that stretches between the brane and the anti-brane. It has charge $+1$ under the brane gauge field and $-1$ under the anti-brane gauge field. Hence it really is charged under a linear combination of the two $U(1)$s that we denote $U(1)_-$ with gauge field $A^- = A^1 -A^2$.

To get the exact coupling between the tachyon and the inflaton one must apply T-duality to this action. 
T-duality acts on the worldsheet states by replacing $\tilde\alpha$ by $-\tilde\alpha$ (and correspondingly for the fermionic modes).  One can work out how the spacetime fields behave under such a transformation (for a complete treatment see \cite{Giveon:1994fu}).  The tachyon is inert since it is the ground state and it has no oscillator dependence. The gauge fields on the other hand transform into scalar fields $A_I \rightarrow \Phi^I/\lambda$. We use greek letters to denote worldvolume coordinates and capitalized latin letters for the transverse coordinates.  The distance between the brane and the anti-brane is a vector $\varphi^I = \Phi^{(1)I} - \Phi^{(2)I}$ and it is T-dual to $A^-$ (we also denote $\Phi^{(1)} = \Phi_{D3}$ and $\Phi^{(2)} = \Phi_{\overline{D3}}$). We use capital $\Phi$ here to distinguish between the canonically normalized fields used in the text.

From the form of the tachyon action (\ref{tachSen}), one can easily see that the interesting coupling we are after, $\varphi^2 T^2$, is coming in the square root term in the covariant derivative of the tachyon. Since T is inert, the potential V(T) remains unchanged as one goes from the $D9\overline{D9}$ down to lower dimensional branes. In particular it does not depend on $\varphi$. Nevertheless, the determinant splits into two parts after the T-duality transformation, a determinant over wordvolume fields and a determinant over transverse fields \cite{Myers:1999ps}. We can factorize the second determinant and incorporate it inside the potential. 
One can calculate the general expression for the action \cite{Garousi:2004rd} but for the purpose of this paper we give the answer for a block diagonal $E_{\mu I} = 0$, no gauge field $A=0$ and a real tachyon $T = \bar T$.
\begin{align}
S &= - \tau_p\int d^{p+1}x \; V(T)V^\prime(T, \varphi^I) \left(\sum_{i=1}^2 \sqrt{-\det\; {\bf A}^{(i)}}\right)\; +\text{C.S.},
\end{align}
where
\begin{align*}
{\bf A}^{(i)} & =  E_{\mu\nu} + \p_\mu \Phi^{(i),K}\p_\nu \Phi^{(i),J} E_{KJ} + \frac{1}{\det Q} 
\p_\mu T\p_\nu T\; ,\\
Q^I_J & = \delta^I_J + \frac{ \varphi^I\varphi^K T^2 E_{KJ}}{\lambda^2}, \\
V^\prime(T, \varphi^I) & = \sqrt{\det Q^I_J}\; .
 \end{align*}
If one expands $V(T)V^\prime(T,\varphi^I)$, one finds the correct mass term as expected from perturbative string theory:
\begin{align}\label{massterm}
V(T)V^\prime(T,\varphi^I) \approx 1 + \frac{T^2}{2} \left( -\frac{1}{2\ap} + \frac{\varphi^I\varphi^JE_{IJ}}{4\pi^2\ap^2}\right)
\end{align}
The Chern-Simons term can be evaluated, it is
\begin{align}
\text{C.S.} = -\tau_p g_s \int V(T)(P[C_{p+1} (\Phi_{D3})] - P[C_{p+1} (\Phi_{\overline{D3}})])
\end{align}
where the coupling to lower dimensional forms are zero for a vanishing gauge field, and for a real tachyon. 

\subsection{Action in the KS Background and Comparison to Inflaton Potential}

We use the metric (\ref{10dmetric}) and turn on only one scalar field in the radial direction of the AdS.  We denote by $\Phi^{(r)}_{D3}$ the position of the D3-brane and $\Phi^{(r)}_{\overline{D}3}$ is the position of the anti-brane ($\varphi$ their difference as before).  For this metric (and with $B_{rr} = 0$),
\begin{align*}
\det Q = 1 + \frac{\varphi^2 T^2 h^{-2}}{\lambda^2}
\end{align*}
and the action reduces to:
\begin{align}
S = &-\tau_3 \int d^4 x V(T) h^4 a^3 \sqrt{1 +  \frac{\varphi^2T^2 h^{-2}}{\lambda^2}} \left( \sqrt{ 1 - \dot\Phi^{(r)2}_{D3} h^{-4} - \frac{\lambda^2h^{-2} \dot T^2}{\lambda^2+ \varphi^2 T^2 h^{-2}}} \right) \\
&+ \tau_3 \int a^3 V(T)h^{4}\nonumber + (\Phi^{(r)}_{D3} \rightarrow \Phi^{(r)}_{\overline{D}3})\nonumber
\end{align}
where the last term is short-hand for the action repeated for the anti-brane. Note that in this expression, the warp factor needs to be evaluated at the position of the D3-brane and the anti-D3-brane respectively, so $h = h(\Phi_{D3}) \approx h(r)$ and $h(\Phi_{\overline{D}3}) = h_A$. Let us check that this has the correct asymptotic behavior for large $\varphi$.  Since the anti-brane sits naturally at the bottom of the throat $\Phi_{\overline{D3}} = \Phi_A$ and $\dot\Phi_{\overline{D3}} = 0$, and $\dot\Phi_{D3} = \dot \varphi$. Both the kinetic and Chern-Simons terms for the anti-brane contribute a constant. Now, far from the tip we can use the metric (\ref{10dmetric}) and $T=0$ and $V(T) = 1$, so that
 \begin{align}
 \label{largedistTach}
 S = -\tau_3\: \int d^4x a^3 \left(h^4\sqrt{1- \dot\varphi^2 h^{-4}} + 2h_A^4 - h^4\right)
 \end{align}

Note that the evaluation of the Chern-Simons term relies on the KS-type background, where only $C_4$ contributes, and the relationship between $C_4$ and the warp factor is imposed by the supergravity equations. This is similar to (\ref{DBIact}), up to the normalization $\dot{\phi}=\dot{\varphi}/\sqrt{\tau_3}$. The other difference comes in the potential where $V_0= 2\tau_3h_A^4$ but the $\phi^2$ term and the coulombic part are missing. For $\varphi > \sqrt{\ap}$ we should expect that quantum effects in the open string theory are important. Indeed one can calculate the open string 1-loop diagram (the cylinder amplitude) to find the coulombic piece of the potential. Alternatively one can account for the coulombic part by calculating the tree level exchange of closed strings (also the cylinder amplitude).  In both cases the results are the same, and one generates a coulombic potential for $\varphi$ \cite{Jones:2002si}. A third method \cite{Kachru:2003sx} is to calculate the perturbation of the background metric due to an extra brane, put the perturbed metric into the brane action, and calculate the potential. This process recovers the coulombic piece and gives the same origin for the constant piece (from the $\overline{D3}$ action) as did the reduction that gave Eq.(\ref{largedistTach}). The result above is to be expected if we take Sen's ``open-string completeness conjecture" \cite{Sen:2003mv, Sen:2004nf} seriously. If we ask what the appropriate action is as the brane moves down the throat, then far from the anti-brane Eq.(\ref{DBIact}) is valid. But as the branes approach within a string length, more and more closed string modes contribute significantly. Rather than calculating all of those contributions, it is much simpler to switch to the action that contains the tachyon field, and which we know correctly reproduces the end point of the brane annihilation. Furthermore, the action gives the expected answer when the branes are far apart, but one must compensate for the disappearance of the tachyon field by including the corrections from closed string exchange.  We expect the coulombic potential to be modified around distances of string length (see \cite{Sarangi:2003sg} for a discussion of this) and then smoothly goes to zero.

The quadratic term for $V(\varphi)$, as was mentioned already, comes from various interactions and quantum effects that are not included in our calculation above. This piece must be added by hand. This is reasonable since the mass term is not related to the brane anti-brane interaction. We expect that this term will depend on the position of the brane, but not the brane anti-brane separation. Therefore the complete correct action is reproduced once quantum corrections are included and the $m^2\phi^2/2$ term is added. 

At small $\varphi$, open string analysis is trustable (for $\varphi$) and we do not have a coulombic quantum correction.  On the other hand we have a tachyon. One might wonder if the mass term for $\varphi$ remains unchanged. We do not see any reasons to expect any significant changes but more work is needed here to completely elucidate this question. This is further addressed in the last section of this appendix.
Putting the brane separation exactly equal to zero, one gets :
\begin{align}
S = -2\tau_3\int d^4x V(T) h_A^4 a^3\sqrt{1- h_A^{-2} \dot T^2} 
\end{align}
This reproduces (\ref{tachact}).

\subsection{Small $\varphi$ Action} 

We are interested in the limit where $\varphi$ is small but not zero.  We also want to keep track of at least one of the angular direction.  Since $\varphi$ is small the AdS approximation is no longer sufficient \cite{Klebanov:2000hb}. The full metric close to the tip is just $\mathbb{R}^{3,1}\times\mathbb{R}^3\times S^3$:
\begin{align}
ds^2 = h_A^2 g_{\mu\nu}dx^\mu dx^\nu + h_A^{-2}\left(\frac{\epsilon^{4/3}}{2^{5/3}3^{1/3}}d\tau^2 + d\Omega^2_3 + \tau^2d\Omega^2_2\right)
\end{align}
Where $\epsilon = h_A^{3/2} 2^{1/4}a_0^{3/8} (g_s M\ap)^{3/4}$ is the deformation parameter and $a_0 = 0.71805$.  At the tip of the throat, $\tau=0$ and the $S^2$ shrinks to zero size. 
We refer to $\tau$ as the radial direction, since it can be related to the usual radial coordinate by: 
\be
\label{rtaurelate}
r-r_{0} = \frac{1}{2^{5/6}3^{1/6}}\epsilon^{2/3} \tau \left(1 + \frac{\tau^{2}}{18} + ... \right)
\ee
The basic point can be illustrated by considering only a single component other than $\tau$. Choosing the longitudinal angle $\psi$ of the $S^3$, the metric is
\begin{align}
ds^2 =  h_A^2 g_{\mu\nu}dx^\mu dx^\nu + h_A^{-2}\epsilon^{4/3} (2/3)^{1/3}\left(\frac{1}{4}d\tau^2 + d\psi^2 + \cdots\right)
\end{align}
Here $\psi$ is the usual azimuthal coordinate in an $S^3$ ranging from $0$ to $\pi$, and one can verify that the radius of the $S^3$ ($R_{S^3}^2 = \epsilon^{4/3} (2/3)^{1/3} = h_A^2 2^{2/3} 3^{-1/3} a_o^{1/2} (g_s M\ap) $) is larger than the warped string length when $g_s M > 1$ .  Note that $\tau$ and $\psi$ are dimensionless. We define fields with dimension of length $\varphi^\tau = R_{S^3}\tau/2$ and $\varphi^\psi = R_{S^3} \psi$. The factor of two is for convenience. 

For small distances the warp factor is essentially constant $h \sim h_A$. With $\dot \Phi_{\overline{D3}} \approx 0$ the action is
\begin{align}
S = &- \int d^4 x V(T)V^\prime(T, \varphi^I)\tau_3h_A^4 a^3 \left(\sqrt{1 - \dot\varphi^I\dot\varphi^JE_{IJ}h_A^{-2} - h_A^{-2} \dot T^2} \right.\\
& \left. + \sqrt{1 - h_A^{-2} \dot T^2}\right) + V(\Phi^{\tau})\nonumber
\end{align}
where
\begin{align}
V^\prime(T,\varphi^I) & = \sqrt{1 +  \frac{\varphi^I\varphi^JE_{IJ}T^2}{\lambda^2}}\; .
\end{align}
We assume that the quadratic piece in the potential $V(\Phi^{\tau})$ only depends the radial part of the brane position, while the angular part is protected by symmetry.

\begin{figure}[htb]
 \begin{center}
\includegraphics[width=0.4\textwidth,angle=0]{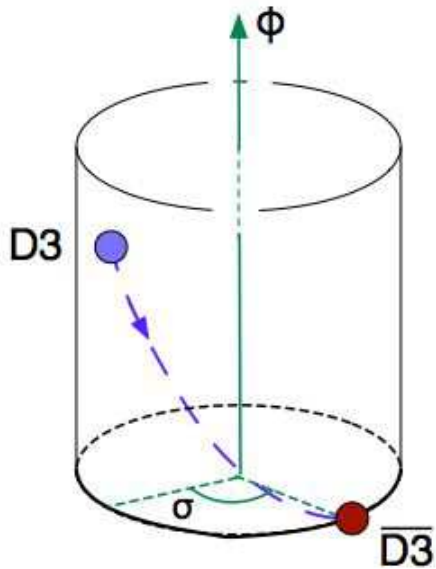} 
\caption{A schematic illustration of the brane position on a simple space $\mathbb{R}^{1}\times S^1$. The anti-brane prefers energetically to sit at $\tau=0$, while the brane moves toward it along the surface of the cylinder (along the dashed blue line connecting the branes). In calculating the brane/anti-brane separation the angular position of the brane is at least as important as the radial position, since the radius (and therefore the circumference) of the $S^1$ is larger than the red-shifted string length.
\label{cylinder}}
\end{center}
\end{figure}

The position fields should be canonically normalized (at $T=0$, and assuming $\varphi^I \approx \Phi^I_{D3}$):  
\ba
\phi&=&\varphi^{\tau} \sqrt{\tau_3} + \phi_A\\\nonumber
\sigma&=&\varphi^{\psi}\sqrt{\tau_3}
\ea
Note that we added $\phi_A$, which corresponds to $r_0$ in (\ref{rtaurelate}). The action is
\begin{align}
S &= - \int d^4 x V(T)V^\prime(T,\phi,\sigma)\tau_3h_A^4 a^3 \left(\sqrt{1 - \dot\phi^2h_A^{-4}\tau_3^{-1} - \dot\sigma^2h_A^{-4}\tau_3^{-1} - h_A^{-2} \dot T^2} \right.\\
 &\left. + \sqrt{1 - h_A^{-2} \dot T^2}\right) + V(\phi,\sigma)\nonumber \, ,
\end{align}
where 
\begin{align}
V(T)V^\prime(T,\phi, \sigma) = \frac{1}{\cosh(\sqrt{\frac{1}{2\ap}}T)}\sqrt{1+ \frac{((\phi-\phi_A)^2 +\sigma^2) h_A^{-2}\tau_3^{-1}T^2}{(2\pi\ap)^2}}\, ,
\end{align}
and $V(\phi,\sigma)$ is the quadratic piece. This is indeed the standard expected form, and examining the expansion of $V^{\prime}(T,\varphi^I)$ establishes that the brane separation depends essentially on $\sigma$ (the angular position on the $S^3$) at the end of the throat. This may be illustrated by picturing the brane moving on a cylinder (the subspace $\mathbb{R}^{1}\times S^1$ at the bottom of the throat) whose radius is larger than the warped string length. This is shown in Figure \ref{cylinder}.

\subsection{The Potential}

A key point for the results presented here is that the brane separation is not the primary inflaton. To keep the problem simple, we also assume that some components of the brane position are partially protected by symmetries, making them effectively irrelevant during most of inflation. These claims are based on several technical points, and there is some difficulty in knowing the exact inflaton trajectory in the six-dimensional space, related to the difficulty in determining the mass term exactly. We must address the contributions to the inflaton mass from the K\"{a}hler potential and the non-perturbative superpotential, as well as compactification effects. 

To begin, suppose that we have simply a D3-brane moving in the conifold. This is a non-compact Calabi-Yau space that has $SU(2)\times SU(2)$ isometry.  In such a space the D3-brane position is a modulus with a flat potential. Upon embedding this cone in a full flux stabilized compactification, we expect to generate a mass term for the position of the D3-brane \cite{Kachru:2003sx}. This term comes from mixing between the position of the D3-brane and the K\"{a}hler moduli, and from distortions of the background geometry due to the brane position, which feed into non-perturbative contributions to the mass \cite{Baumann:2006th}. This last effect was originally computed as quantum corrections to the K\"{a}hler potential \cite{Berg:2004ek} but was reproduced in a much simpler and enlightening calculation in \cite{Baumann:2006th}. Because this picture gives a cleanly geometric view of the effect, it is clear that it is the position of the brane relative to certain 4-cycles, and not relative to the anti-brane, that enters. Also, one expects that the anti-brane in no way cancels the effect, since it provides an energy density that would similarly distort the cycles. 

A geometric picture for the K\"{a}hler potential is less clear in realistic cases. In general, one might expect some nearly flat directions simply from the fact that fields (certainly the K\"{a}hler and complex structure moduli and probably the brane position) naturally enter the K\"{a}hler potential as $(\phi-\bar{\phi})$. It is this term that we expect to generate a large mass for the radial direction, but not for the angles. A simple example supporting this picture is considered in \cite{Angelantonj:2003zx}. However, it is well known that compact Calabi-Yau spaces do not have exact isometries, so the discussion for simple spaces may be misleading. Discussions of these issues can be found in \cite{Shandera:2004zy}. This question is also helpfully addressed in the appendix of \cite{DeWolfe:2004qx}, where the effective masses generated by compactifying and breaking the isometries of the conifold are estimated. The mass term generated for all components in this way is suppressed relative to other effects by the warping at the tip of the throat. This can be intuitively understood since the process of compactification involves more distant, bulk effects, while the fluxes involved in moduli stabilization for the throat (and so the primary mass term) are local and preserve the isometries. 

In summary, it is difficult to give a simple expression for the inflationary trajectory in terms of the brane coordinates because there are many contributions to the quadratic piece of the potential. However, it seems quite reasonable that the combination of the brane position moduli that enters into the mass term is different from the $\DD$ separation, which is all that is needed for additional density perturbations to be generated. If some directions (e.g. the angles) are not somewhat protected, then the problem may become a true multi-field model which is messier to deal with, and would almost certainly make different predictions than those calculated so far. In that sense, our analysis here may be only the tip of the iceberg.

\bibliographystyle{JHEP}

\providecommand{\href}[2]{#2}\begingroup\raggedright\endgroup

\end{document}